\documentclass[a4paper]{article}

\usepackage{INTERSPEECH2021}
\usepackage{multirow}
\usepackage{booktabs}
\usepackage{ulem}
\usepackage[dvipsnames]{xcolor}
\newcommand\hl{\bgroup\markoverwith
  {\textcolor{yellow}{\rule[-.5ex]{2pt}{2.5ex}}}\ULon}

\title{Restoring degraded speech via a modified diffusion model}
\name{Jianwei Zhang$^1$, Suren Jayasuriya$^1$, Visar Berisha$^1$}
\address{
  $^1$Arizona State University}
\email{\{jzhan396,sjayasur,visar\}@asu.edu}

\begin{document}

\maketitle
\begin{abstract}

There are many deterministic mathematical operations (e.g. compression, clipping, downsampling) that degrade speech quality considerably. In this paper we introduce a neural network architecture, based on a modification of the DiffWave model, that aims to restore the original speech signal. DiffWave, a recently published diffusion-based vocoder, has shown state-of-the-art synthesized speech quality and relatively shorter waveform generation times, with only a small set of parameters. We replace the mel-spectrum upsampler in DiffWave with a deep CNN upsampler, which is trained to alter the degraded speech mel-spectrum to match that of the original speech. The model is trained using the original speech waveform, but conditioned on the degraded speech mel-spectrum. Post-training, only the degraded mel-spectrum is used as input and the model generates an estimate of the original speech. Our model results in improved speech quality (original DiffWave model as baseline) on several different experiments. These include improving the quality of speech degraded by LPC-10 compression, AMR-NB compression, and signal clipping. Compared to the original DiffWave architecture, our scheme achieves better performance on several objective perceptual metrics and in subjective comparisons. Improvements over baseline are further amplified in a out-of-corpus evaluation setting.

\end{abstract}

\section{Introduction}

Many algorithms and mathematical operations degrade the quality of speech. For example, speech compression algorithms reduce the sampling rate and use linear predictive coding to compress the input; speech-clipping introduces high-frequency content with a negative impact on quality. Reduced speech quality can impact intelligibility and makes the resulting speech less suitable for downstream applications like automatic speech recognition or speaker identification algorithms. Speech enhancement (SE) of degraded speech is important across many applications including telecommunications~\cite{tan2019real}, speech recognition~\cite{subramanian2019speech}, etc. Many methods have been developed for similar applications, such as speech denoising and dereverberation~\cite{lan2020combining,su2020hifi}. Most current SE methods are designed to remove background noise, most are additive noise models. However, the aforementioned operations (e.g. compression, clipping) are non-linear and lossy. The goal of this paper is to restore the degraded speech generated by lossy deterministic transformations.

Broadly speaking, there are two families of SE techniques: based on traditional statistical signal processing and based on machine learning. Traditional methods include statistical model-based methods~\cite{loizou2013speech}, e.g.  spectral subtraction~\cite{hasan2004modified}, and Wiener filtering~\cite{abd2008speech}. While these methods work well for additive noise conditions, they are not suitable for our application. Enhancement methods based on machine learning models such as diffusion models and U-nets with adversarial loss have resulted in a sizeable improvement in performance~\cite{tan2019real,lan2020combining,su2020hifi,pascual2017segan}. These models can enhance speech quality, however they require complex network structures with a large number of parameters. Our aim is to develop sample-efficient networks trained to invert the lossy transformation and impute the missing information in the signal. We posit that for deterministic transformations (e.g. compression, clipping), we can efficiently learn the inversion and generate high-quality speech by leveraging state-of-art vocoders.

Modern vocoders can generate high-quality speech based on an input conditioner, e.g. a mel-spectrum. A widely used ML-based vocoder is WaveNet~\cite{oord2016wavenet}. It can synthesize high-quality speech, but the synthesis run-time is slow. WaveFlow is a flow-based ML vocoder with short generation time, however it contains a large number of parameters~\cite{ping2020waveflow}. Many vocoders (e.g. LPCNet) have recently focused on improving synthesis efficiency~\cite{valin2019lpcnet}. DiffWave, a diffusion model-based vocoder was recently published with state-of-the-art synthesized speech quality, a relatively short waveform generation time, and small number of parameters~\cite{kong2020diffwave}. However, DiffWave was primarily used for generative modeling tasks such as unsupervised speech generation where the data distribution of audio was learned by the model.

In this paper, our key insight is that a diffusion-based model such as DiffWave can be trained in a supervised fashion to restore degraded speech, particularly for these deterministic operations. To do so, we condition DiffWave on the degraded mel-spectrum of the input speech, and train the network to recover back the original speech. However, we observe that this method only achieves partial recovery of the original speech. To further improve performance, we modify DiffWave network architecture by including a pre-trained inversion network to restore the quality and intelligibility of speech. We replace the upsampling layers in a pre-trained DiffWave model with a deep CNN upsampler, which has the capacity to learn an inversion model that alters the degraded speech mel-spectrum to generate the conditioner for restored speech synthesis by DiffWave model.  

To validate our methods, we perform experiments to compare the quality and intelligibility of restored audio when degraded by three deterministic lossy mathematical operations: linear predictive coding (LPC-10) compression, adaptive multi-rate narrow-band (AMR-NB) compression, and signal clipping. We compare the results of the original DiffWave trained in a supervised fashion as well as our modified DiffWave model with inversion module. Our results show that our modified model improves on the original DiffWave model for this application, restoring speech quality and intelligibility on both in-corpus (out-of-sample) and cross-corpus evaluations. In summary, our contributions are: 1) We demonstrate that DiffWave is able to produce better-quality speech, even conditioned on a distorted mel-spectrum, and 2) we modify DiffWave’s architecture with a deep CNN upsampling network for the conditioner, resulting in superior quality in speech restoration.

\section{Methods}

In this section, we describe our network architecture and training approach. We first describe how the original DiffWave model can be trained to restore degraded speech - this serves as our baseline model. We then describe our modifications to the DiffWave vocoder using a deep CNN inversion network to further enhance performance.

\subsection{DiffWave for restoring degraded speech}

DiffWave is a speech waveform generative model, i.e. a vocoder, based on diffusion models~\cite{kong2020diffwave}. It takes the mel-spectrum as conditioning input and generates corresponding speech as shown in the top part of Figure~\ref{fig:structure}. DiffWave was not originally designed for SE, but we use it for restoring lossy transformed speech in this paper. We train the DiffWave vocoder by using paired original speech $x$ and degraded speech mel-spectrum $m_{T}$ samples (in original paper, the clean mel-spectrum was used). Once the model converges, we use it to generate the estimated original speech $\hat{x'}$ conditioning on corresponding degraded speech mel-spectrum $m_{T}$. In the experimental results section, we show that a supervised DiffWave can restore the quality to a certain extent. After analyzing the structure of DiffWave, we identified the upsampler as a key component that can be further optimized to improve quality.

\subsection{Deep CNN for Conditioner Upsampling}

\begin{figure}[t]
  \centering
  \includegraphics[width=\linewidth]{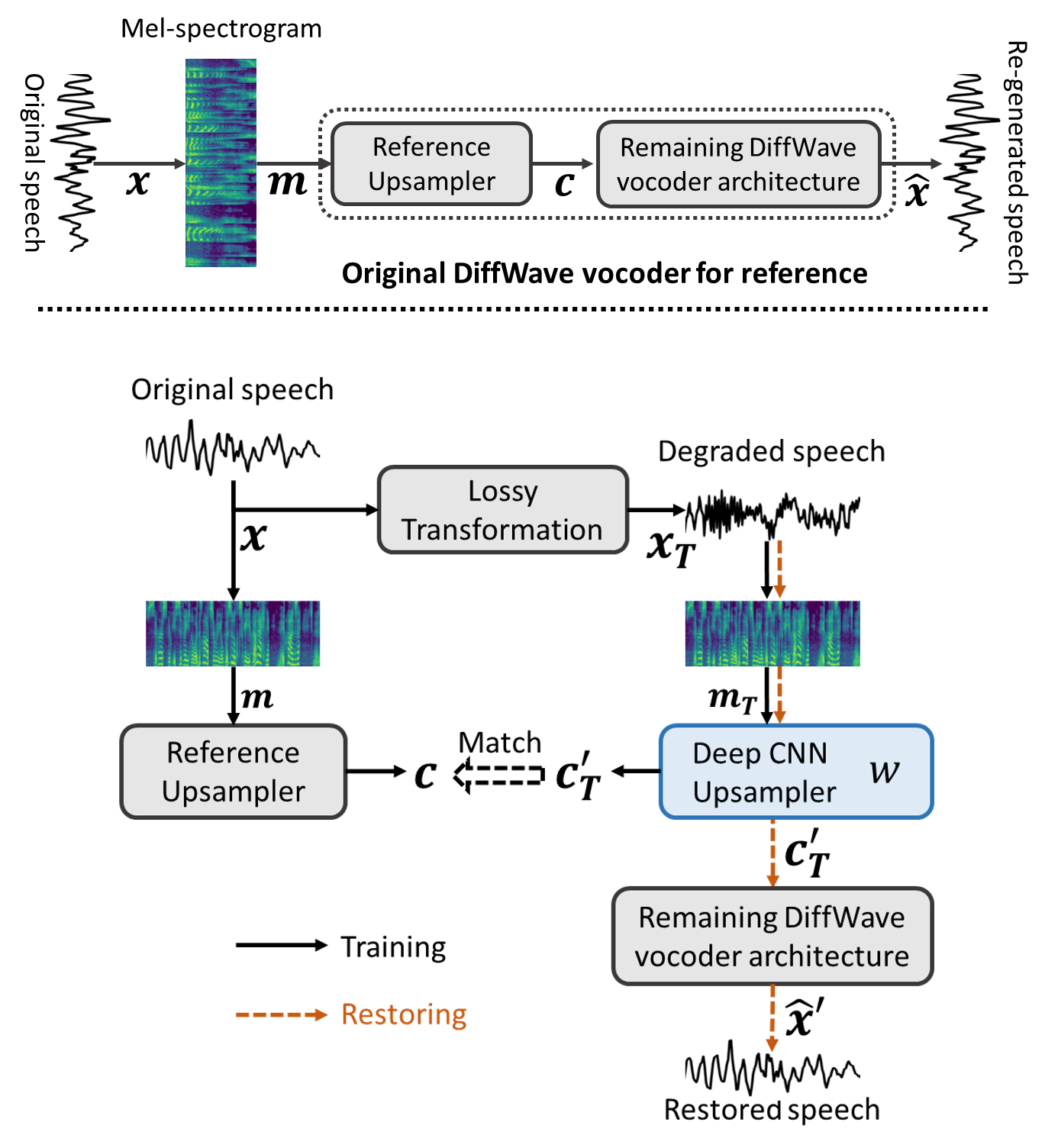}
  \caption{Top: Supervised training for the original DiffWave model. Bottom: Our method for training a deep CNN upsampler to match the conditioner of DiffWave's reference upsampler. Then the remaining DiffWave vocoder architecture is used for restored speech waveform generation.}
  \label{fig:structure}
\end{figure}

The DiffWave model contains three modules: an upsampler network, a diffusion embedding network, and residual learning blocks (for network details, please refer to the original paper~\cite{kong2020diffwave}). In Diffwave, the upsampler network is used to increase the dimension of the input mel-spectrum to be the conditioner for speech waveform synthesis. The structure of the upsampler in the original DiffWave model is simple, it contains two 2D convolutional transposed layers.

We propose a new upsampler network, i.e. a deep CNN upsampler, to replace the original one. The degraded speech mel-spectrum $m_{T}$ passes through several CNN nets with increasing channel size. The increased capacity of the upsampler allows us to invert the non-linear transformation and impute the lost information. This is then fed through cross-stacked CNN layers and transpose layers to decrease the channel size while increasing the mel-spectrum dimension to match the output speech waveform's dimension.

In our experiments, we found that simply replacing DiffWave's upsampler with our own network did not result in improved performance. The training of a diffusion-model with the CNN upsampler led to poor convergence to a local minima similar to training the original DiffWave.

To overcome this, we found it was better to separately train the CNN upsampler, independent of DiffWave, but with the criterion to match DiffWave's upsampling network's output on the original speech. In this scheme, we first train the DiffWave vocoder model which maps $x \rightarrow \hat{x}$, i.e. we train the model to generate an estimated original speech waveform conditioned on the original speech mel-spectrum. As shown in bottom part of Figure~\ref{fig:structure}, DiffWave's upsampler is then extracted as the reference upsampler for the deep CNN upsampler training. The remaining DiffWave vocoder architecture is used for restored speech waveform synthesis. To train the deep CNN upsampler, we first generate the reference conditioner $c$ from original speech mel-spectrum $m$ via a reference upsampler, and generate the altered conditioner $c'_{T}$ from the corresponding degraded speech mel-spectrum $m_{T}$ with our new upsampler. We train the new upsampler with a mean absolute error loss (L1 loss),

\begin{equation}
    \ell (c_{n},c'_{T_n};w) = \frac{1}{N} \sum_{n=1}^{N} |c_{n}-c'_{T_n}|,
\end{equation}

\noindent where $c'_{T_n}$ is given by deep CNN upsampler with weights $w$. After training the upsampler, at inference we simply feed the degraded speech mel-spectrum $m_{T}$ through the new deep CNN upsampler to generate altered conditioner $c'_{T}$, and then through remaining DiffWave vocoder architecture to generated the estimated original speech $\hat{x'}$.

\section{Experiments}

\begin{table*}[ht]
\footnotesize
\centering
\caption{Quantitative measures of speech quality for in-corpus and cross-corpus evaluations. The comparisons are between the baseline model ('DW'), the modified DiffWave architecture ('ModDW'), and input degraded speech ('Degraded'). Each score is an average from a randomly-selected set of 128 samples, with standard deviation in parentheses. An asterisk means that the difference between ModDW and DW is statistically significant with $p<0.05$ by doing t-test.}
\label{tab:perceptual}
\begin{tabular}{c|c|cccccc}
                                                                                     & \textbf{Tranformation}                                                        & \textbf{Model} & \textbf{PFP Loss}        & \textbf{PESQ}            & \textbf{CSIG}            & \textbf{CBAK}            & \textbf{COVL}            \\ \midrule[1pt]
\multirow{9}{*}{\begin{tabular}[c]{@{}c@{}}In-\\ corpus\\ (TIMIT)\end{tabular}}      & \multirow{3}{*}{\begin{tabular}[c]{@{}c@{}}LPC-10\\ Compression\end{tabular}} & Degraded       & 0.0173(0.0010)           & 1.2029(0.1122)           & 1.9829(0.3419)           & 1.5589(0.1747)           & 1.4826(0.2501)           \\
                                                                                     &                                                                               & DW             & 0.0140(0.0009)           & 1.2401(0.1216)           & 2.7146(0.3138)           & 1.7311(0.1857)           & 1.8833(0.2543)           \\
                                                                                     &                                                                               & ModDW          & \textbf{0.0121(0.0009)*} & \textbf{1.5056(0.2287)*} & \textbf{3.1048(0.2865)*} & \textbf{1.8705(0.1781)*} & \textbf{2.2390(0.2794)*} \\ \cline{2-8} 
                                                                                     & \multirow{3}{*}{\begin{tabular}[c]{@{}c@{}}AMR-NB\\ Compression\end{tabular}} & Degraded       & 0.0150(0.0006)           & 2.2787(0.2937)           & 2.8363(0.4383)           & 2.3645(0.1581)           & 2.5355(0.3529)           \\
                                                                                     &                                                                               & DW             & 0.0130(0.0008)           & 2.0022(0.2661)           & 3.1793(0.2508)           & 2.2444(0.1415)           & 2.5687(0.2506)           \\
                                                                                     &                                                                               & ModDW          & \textbf{0.0112(0.0006)*} & \textbf{2.4498(0.3421)*} & \textbf{3.5618(0.2812)*} & \textbf{2.5127(0.1791)*} & \textbf{3.0008(0.3070)*} \\ \cline{2-8} 
                                                                                     & \multirow{3}{*}{\begin{tabular}[c]{@{}c@{}}Signal\\ Clip (25\%)\end{tabular}} & Degraded       & 0.0116(0.0006)           & 1.5439(0.2155)           & 2.3717(0.2622)           & 1.8279(0.1699)           & 1.7797(0.2211)           \\
                                                                                     &                                                                               & DW             & 0.0112(0.0004)           & 1.5022(0.1859)           & 2.5630(0.2084)           & 1.9280(0.1988)           & 1.8145(0.2632)           \\
                                                                                     &                                                                               & ModDW          & \textbf{0.0096(0.0003)*} & \textbf{2.2144(0.2845)*} & \textbf{2.6871(0.2544)*} & \textbf{2.5410(0.1831)*} & \textbf{2.2687(0.2988)*} \\ \midrule[1pt]
\multirow{9}{*}{\begin{tabular}[c]{@{}c@{}}Cross-\\ corpus\\ (Mozilla)\end{tabular}} & \multirow{3}{*}{\begin{tabular}[c]{@{}c@{}}LPC-10\\ Compression\end{tabular}} & Degraded       & 0.0156(0.0019)           & 1.2134(0.1086)           & 2.0628(0.3548)           & 1.4790(0.1573)           & 1.5254(0.2281)           \\
                                                                                     &                                                                               & DW             & 0.0154(0.0021)           & 1.2088(0.1173)           & 2.5583(0.3408)           & 1.5060(0.2046)           & 1.7547(0.2453)           \\
                                                                                     &                                                                               & ModDW          & \textbf{0.0132(0.0022)*} & \textbf{1.3499(0.2165)*} & \textbf{2.7654(0.3624)*} & \textbf{1.6039(0.2365)*} & \textbf{1.9536(0.2959)*} \\ \cline{2-8} 
                                                                                     & \multirow{3}{*}{\begin{tabular}[c]{@{}c@{}}AMR-NB\\ Compression\end{tabular}} & Degraded       & 0.0145(0.0015)           & 1.7621(0.2779)           & 2.5079(0.4845)           & 2.0105(0.1617)           & 2.0700(0.3487)           \\
                                                                                     &                                                                               & DW             & 0.0144(0.0013)           & 1.6875(0.2447)           & 2.5943(0.4694)           & 1.8846(0.1780)           & 2.0557(0.3276)           \\
                                                                                     &                                                                               & ModDW          & \textbf{0.0129(0.0011)*} & \textbf{1.8793(0.3267)*} & \textbf{2.8109(0.4629)*} & \textbf{2.0763(0.1921)*} & \textbf{2.2853(0.3698)*} \\ \cline{2-8} 
                                                                                     & \multirow{3}{*}{\begin{tabular}[c]{@{}c@{}}Signal\\ Clip (25\%)\end{tabular}} & Degraded       & 0.0120(0.0007)           & 1.3540(0.1569)           & 2.9644(0.3566)           & 1.5659(0.0994)           & 2.1180(0.2467)           \\
                                                                                     &                                                                               & DW             & 0.0122(0.0008)           & 1.2156(0.1355)           & 3.0144(0.3266)           & 1.6804(0.1698)           & 2.0098(0.2554)           \\
                                                                                     &                                                                               & ModDW          & \textbf{0.0115(0.0005)*} & \textbf{2.0756(0.4285)*} & \textbf{3.4742(0.4177)*} & \textbf{2.2240(0.2213)*} & \textbf{2.7385(0.4074)*} \\ \midrule[1pt]
\end{tabular}
\end{table*}

\subsection{Implementation Details}

\begin{figure}[t]
  \centering
  \includegraphics[width=\linewidth]{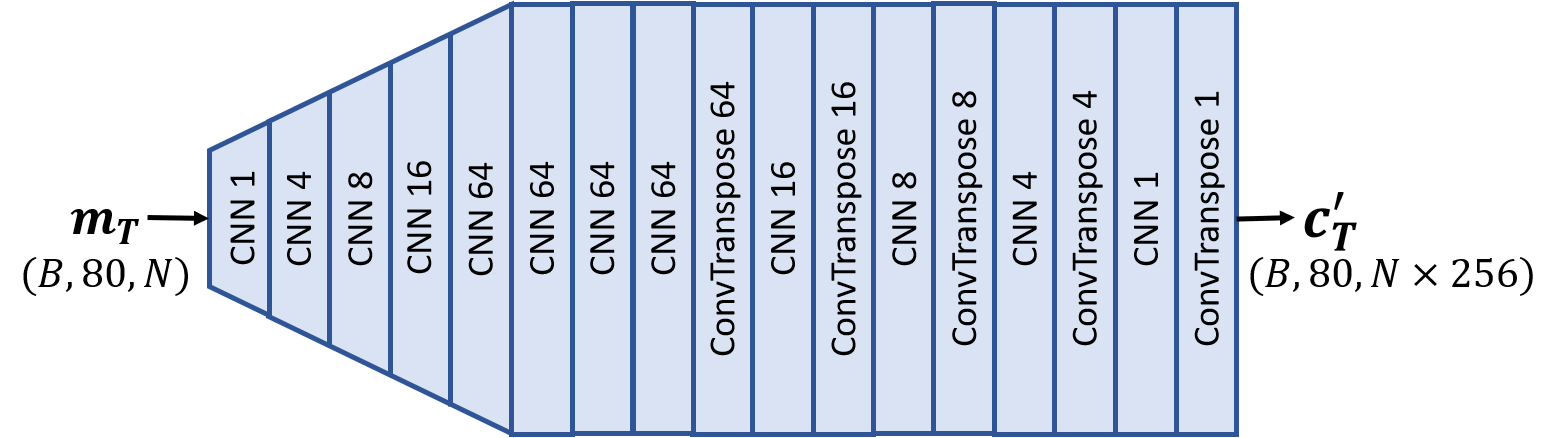}
  \caption{The network structure of our deep CNN upsampler.}
  \label{fig:net}
\end{figure}

\noindent \textbf{Network Architecture:} Our network consists of a 15-layer CNN with a largest channel size of 64, as shown in Figure~\ref{fig:net}. The first 8 layers are 2-D CNNs: kernal size of (5,5) and stride of (1,1) across the layers; channel size of 1, 4, 8, 16, 64, 64, 64, 64; each layer is stacked with a 2-D batch normalization and a leaky-relu whose negative slope is 0.4. The next 7 layers is a cross-stacked 2-D convolutional transpose net and 2-D CNN. For the 2-D convolutional transpose net, the kernel size is (3,8), stride size is (1,4), and the channel size is kept the same as the input. For the 2-D CNN, the settings are the same and the channel size is 64, 16, 8, 4, 1. Again, each layer is stacked with a 2-D batch normalization and a leaky-relu whose negative slope is 0.4. These settings ensure the generated conditioner from the deep CNN upsampler has the same dimensions as that generated by the reference upsampler. This network architecture seemed to provide a good balance on the trade-off between model performance and the size of model parameters set, and we performed ablation studies on the layer sizes and dimensions to arrive at this final architecture.

As discussed in Section 2.2, training happens in two stages. First, we train the DiffWave vocoder from a PyTorch implementation~\cite{lmntdiffwave2020}, i.e. train the model to generate the original speech waveform conditioning on the original speech's mel-spectrum. We used the TIMIT~\cite{garofolo1993darpa} training dataset, a widely used English speech dataset, for training. The DiffWave vocoder was trained for 1M steps (100 hours on 2 Titan Xp GPUs) with a learning rate of 0.0002. For the second stage of training, the deep CNN upsamper is trained to alter the upsampled conditioner from the degraded speech mel-spectrum to match that generated by the reference upsampler from the paired original speech mel-spectrum. We train the upsampler for approximately 50k steps (6 hours on 1 Titan Xp GPU) with a learning rate of 0.001. We use the Adam optimizer for all trainings~\cite{kingma2014adam}.

\noindent\textbf{Lossy Operations:} In this paper, we conduct three experiments to evaluate our model: 1) Restoring speech compressed by the LPC-10 algorithm~\cite{tremain1982government}, 2) Restoring speech compressed by the AMR-NB algorithm (mode: MR515, bit rate = 5.15 kbit/s)~\cite{amrref}, and 3) Restoring speech with clipped magnitude (25\% of the highest-energy samples clipped).

\noindent\textbf{Datasets:} For all three experiments, we use the TIMIT training and testing dataset as our training and in-corpus evaluation dataset correspondingly. The speech in TIMIT is regarded as original speech, and we use the three different algorithms to generate degraded speech files. We also conduct a cross-corpus evaluation for each of the three conditions. We use the Mozilla common voice English dataset for the cross-corpus evaluation \cite{commonvoice:2020}. This is a large corpus that contains more than 1,500 hours of short sentences read by English speakers with various accents, ages, and genders across the world. We randomly selected 128 speech samples and downsampled to 16 kHz. The cross-corpus evaluation did not involve additional training or fine-tune. Note that all experiments mentioned in the paper are based on 16 kHz speech.

\noindent\textbf{Evaluation metrics:} To evaluate the restored speech quality quantitatively, we choose metrics used widely in speech enhancement, namely PESQ~\cite{rix2001perceptual}, CSIG, CBAK and COVL~\cite{hu2007evaluation}, and the phone-fortified perceptual (PFP) loss described in \cite{hsieh2020improving}. We did not use any of these metrics during training. PESQ, CSIG, CBAK, and COVL have been shown to correlate with ``quality", whereas the PFP loss is a proxy for ``intelligibility" as it is based on a speech recognition model. For all metrics, the required reference signal is the original speech.

\noindent\textbf{Baseline model:} Our baseline model is the original DiffWave model trained for restoring degraded speech as described in Section 2.1. For all three experiments, the DiffWave model is trained with the original speech waveform and corresponding degraded speech mel-spectrum.

\subsection{Objective evaluations}

\begin{figure}[t]
  \centering
  \includegraphics[width=\linewidth]{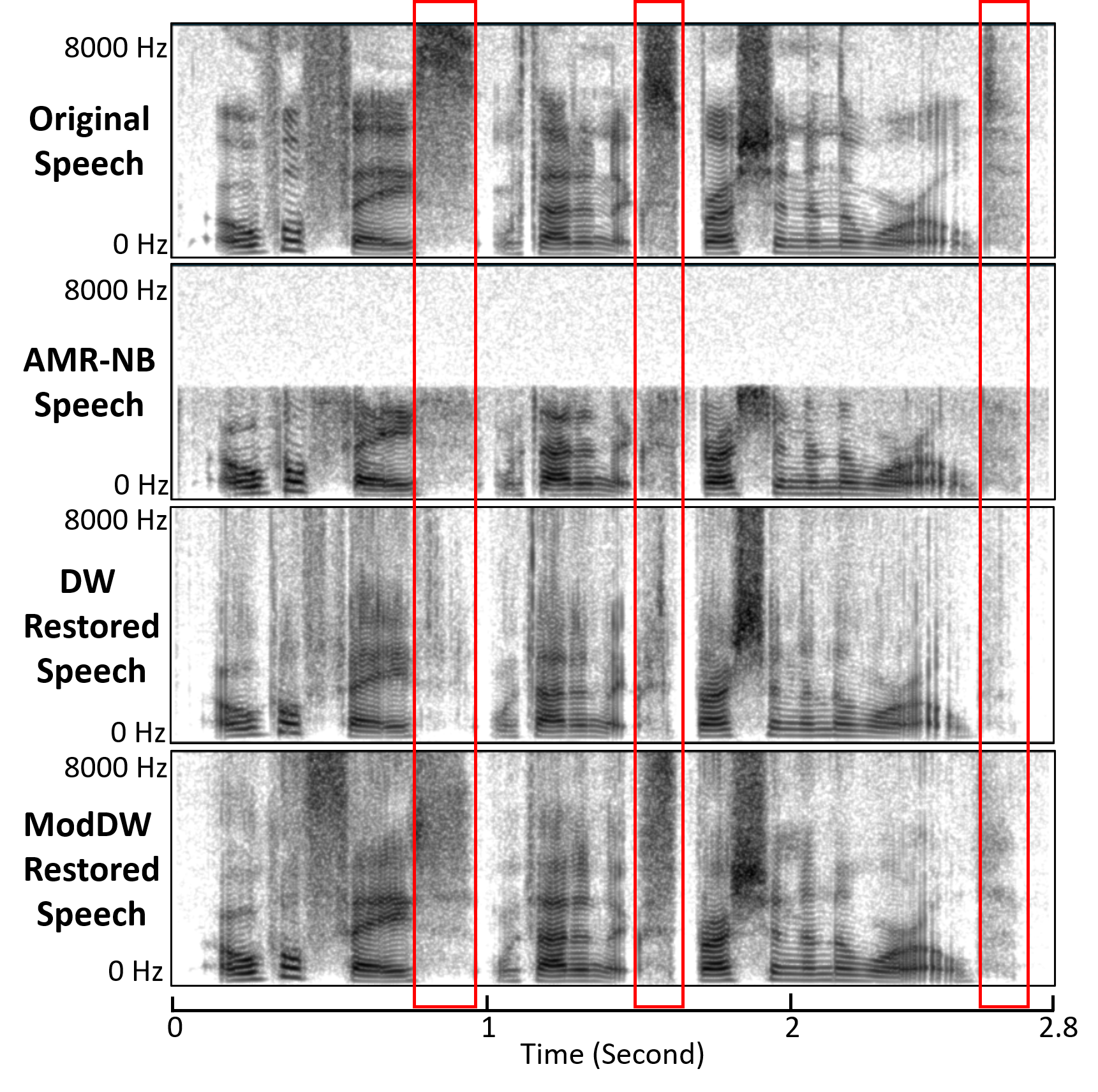}
  \caption{A comparison of spectra between the original speech, degraded speech, baseline model, and modified DiffWave model. Samples are from the AMR-NB experiment on a TIMIT sample. The differences in high-frequency restoration are apparent in the highlighted regions. }
  \label{fig:spec}
\end{figure}

Table \ref{tab:perceptual} shows objective measures for in-corpus and cross-corpus evaluations of the baseline model (labeled as 'DW'), our proposed modified DiffWave scheme (labeled as 'ModDW'), and the input degraded speech (labeled as 'Degraded'). Comparing the score of the three operations, they have varying effects on speech quality. The LPC-10 compressed speech results in the poorest quality speech; whereas the  AMR-NB compressed speech has the highest score on conventional perceptual score but the lowest on PFP loss, which indicates the AMR-NB compressed speech is of higher quality but is less intelligible. The worse PFP scores are likely due to the fact that AMR-NB downsamples the audio to 8 khz, removing all high-frequency content beyond 4 kHz.

Comparing the PFP loss for the baseline model and degraded speech, the baseline can restore degraded speech intelligibility under the in-corpus situation. However, for the conventional perceptual score (e.g. PESQ) it does not show significant improvement, in some cases the quality is poorer than the degraded speech (for AMR-NB, PESQ 2.00 $<$ 2.28). In cross-corpus evaluations, the baseline model failed to restore the degraded speech. The PFP loss for the baseline model is close or higher than the degraded speech. The results indicate that the baseline models fails to generalize outside the training set.

The modified DiffWave model proposed in this paper surpasses the baseline model significantly both for in-corpus and cross-corpus evaluation for all measures. All modified DiffWave model scores are higher than degraded speech, which means our model can restore the quality of different degraded speech sets at evaluation time. In our clipping experiment, the modified DiffWave model achieves a PFP score of 0.0098 in in-corpus evaluation, which nearly matches that of the original speech. Figure~\ref{fig:spec} shows an example spectra from the AMR-NB experiment for the in-corpus evaluation dataset. Our modified model can more accurately impute missing information in the high frequency band relative to the baseline model. It is important to note that the cross-corpus evaluation is especially difficult. This corpus contains sentences recorded by English speakers with various ages, genders, and accents/dialects. This provides strong evidence of generalizability.

\subsection{Subjective evaluations}

\begin{figure}[t]
  \centering
  \includegraphics[width=\linewidth]{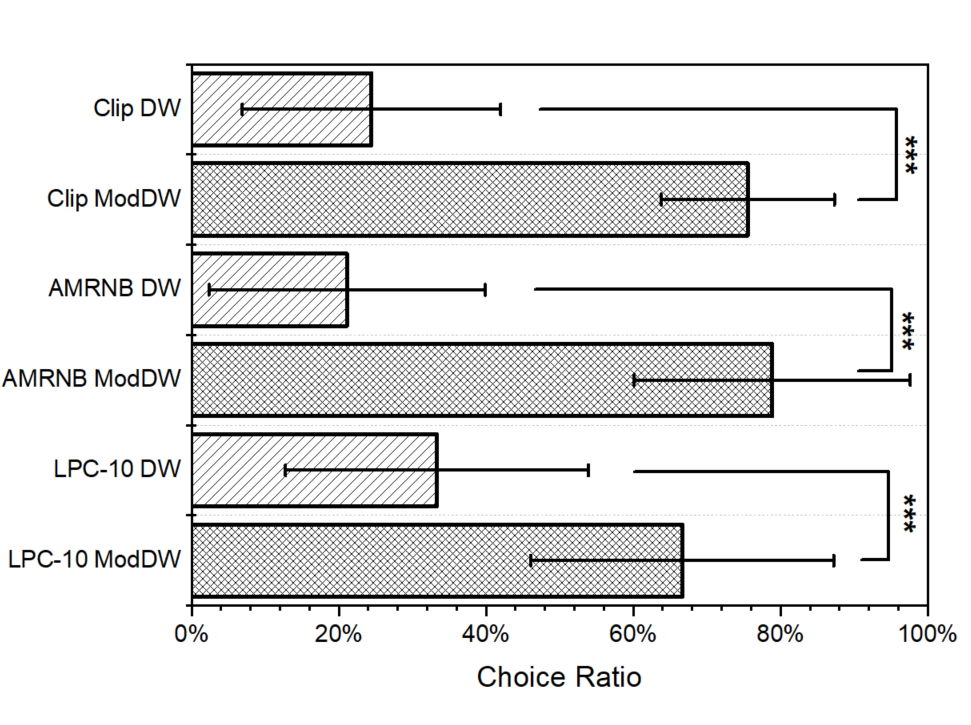}
  \caption{Results of AB preference test for three experiments. $***$ indicates significant difference at $p<0.001$ by doing t-test.}
  \label{fig:abtest}
\end{figure}

The perceptual measures used in this paper are imperfect proxies for human perception, as the restored speech's perceptual measures can be worse but listeners could still think the speech sounds better. We encourage the reader to listen to the speech samples in the supplemental material to determine for themselves the quality of the reconstructed speech.

To compare our methods subjectively, we conducted AB preference test to compare the baseline model with our modified DiffWave model performance on restoring degraded speech. For each listening test, 15 pairs restored speech samples generated randomly from the TIMIT evaluation dataset by original DiffWave model and our modified model (5 pairs from the LPC-10 experiment, 5 pairs from the AMR-NB experiment, 5 pairs from the signal clipping experiment). The order of presentation of the DW-ModDW pair is randomized, and all 18 listeners evaluated a different selection of 15 sentences. We ensure the same spoken sentence is not used twice in any of the pairs. A total of 18 listeners participated in the study, and they were instructed to select the sample with better quality without knowledge of what method generated the sample. The AB preference results are shown in Figure~\ref{fig:abtest}. We can observe that our modified DiffWave model significant outperforms (with p-value $<$ 0.001) the baseline model in all three experiments.

\section{Conclusions}

In this paper, we present a modified DiffWave model for superior quality restoration from distorted and lossy speech. We first train the DiffWave vocoder model to restore degraded speech in supervised fashion and produce good results. We also proposed a modified model that uses a deep CNN upsampler to replace original upsampler in DiffWave. Extensive in-corpus, cross-corpus and subjective perceptual evaluations show that the modified DiffWave model outperforms the original model in restoring degraded speech generated by lossy transformations.

Our modified model can revert the deterministic transformation. Future work will focus on extending this scheme to scenarios where the transformation is stochastic (e.g. noisy speech).

\section{Acknowledgements}

This work was partially supported by ONR Contract N000142012330, and by NIH NIDCD R01 DC006859.

\bibliographystyle{IEEEtran}

\bibliography{mybib}

\end{document}